\def\BZT{{\rm Z{\hbox to 3pt{\hss\rm Z}}}}
\def\BZS{{\hbox{\sevenrm Z{\hbox to 2.3pt{\hss\sevenrm Z}}}}}
\def\BZSS{{\hbox{\fiverm Z{\hbox to 1.8pt{\hss\fiverm Z}}}}}
\def\BZ{{\mathchoice{\BZT}{\BZT}{\BZS}{\BZSS}}}
\def\BQT{\,\hbox{\hbox to -2.8pt{\vrule height 6.5pt width .2pt\hss}\rm Q}}
\def\BQS{\,\hbox{\hbox to -2.1pt{\vrule height 4.5pt width .2pt\hss}$
    \scriptstyle\rm Q$}}
\def\BQSS{\,\hbox{\hbox to -1.8pt{\vrule height 3pt width
    .2pt\hss}$\scriptscriptstyle \rm Q$}}
\def\BQ{{\mathchoice{\BQT}{\BQT}{\BQS}{\BQSS}}}
\def\BCT{\,\hbox{\hbox to -3pt{\vrule height 6.5pt width .2pt\hss}\rm C}}
\def\BCS{\,\hbox{\hbox to -2.2pt{\vrule height 4.5pt width .2pt\hss}$
    \scriptstyle\rm C$}}
\def\BCSS{\,\hbox{\hbox to -2pt{\vrule height 3.3pt width
    .2pt\hss}$\scriptscriptstyle \rm C$}}
\def\BC{{\mathchoice{\BCT}{\BCT}{\BCS}{\BCSS}}}
\def\BHT{{\rm I{\hbox to 5.3pt{\hss\rm H}}}}
\def\BHS{{\hbox{\sevenrm I{\hbox to 4.2pt{\hss\sevenrm H}}}}}
\def\BHSS{{\hbox{\fiverm I{\hbox to 3.5pt{\hss\fiverm H}}}}}

\def\BPT{{\rm I{\hbox to 5pt{\hss\rm P}}}}
\def\BPS{{\hbox{\sevenrm I{\hbox to 4pt{\hss\sevenrm P}}}}}
\def\BPSS{{\hbox{\fiverm I{\hbox to 3pt{\hss\fiverm P}}}}}

\def\BST{\;\hbox{\hbox to -4.5pt{\vrule height 3pt width .2pt\hss}
    \raise 4pt\hbox to -2pt{\vrule height 3pt width .2pt\hss}\rm S}}
\def\BSS{\;\hbox{\hbox to -4.2pt{\vrule height 2.3pt width .2pt\hss}
    \raise 2.5pt\hbox to -4.8pt{\vrule height 2.3pt width .2pt\hss}
    $\scriptstyle\rm S$}}
\def\BSSS{\;\hbox{\hbox to -4.2pt{\vrule height 1.5pt width .2pt\hss}
    \raise 1.8pt\hbox to -4.8pt{\vrule height 1.5pt width .2pt\hss}
    $\scriptscriptstyle\rm S$}}

\def\BFT{{\rm I{\hbox to 5pt{\hss\rm F}}}}
\def\BFS{{\hbox{\sevenrm I{\hbox to 4pt{\hss\sevenrm F}}}}}
\def\BFSS{{\hbox{\fiverm I{\hbox to 3pt{\hss\fiverm F}}}}}

\def\BNT{{\rm I{\hbox to 5pt{\hss\rm N}}}}
\def\BNS{{\hbox{\sevenrm I{\hbox to 4pt{\hss\sevenrm N}}}}}
\def\BNSS{{\hbox{\fiverm I{\hbox to 3pt{\hss\fiverm N}}}}}
\def\BN{{\mathchoice{\BNT}{\BNT}{\BNS}{\BNSS}}}
\def\BRT{{\rm I{\hbox to 5.5pt{\hss\rm R}}}}
\def\BRS{{\hbox{\sevenrm I{\hbox to 4.3pt{\hss\sevenrm R}}}}}
\def\BRSS{{\hbox{\fiverm I{\hbox to 3.35pt{\hss\fiverm R}}}}}
\def\BR{{\mathchoice{\BRT}{\BRT}{\BRS}{\BRSS}}}
\def\BAT{\hbox{\raise1.8pt\hbox{\sevenrm/}{\hbox to 4pt{\hss\rm A}}}}
\def\BAS{\hbox{\raise1.4pt\hbox{\fiverm/}{\hbox to 3pt{\hss\sevenrm A}}}}
\def\BASS{\hbox{\raise1.4pt\hbox{\fiverm/}{\hbox to 3pt{\hss\sevenrm A}}}}

\def\qABF{[1]}
\def\qBax{[28]}
\def\qBaRe{[10]}
\def\qBPZ{[3]}
\def\qBla{[29]}
\def\qBlu{[20]}
\def\qBre{[18]}
\def\qCas{[30]}
\def\qChr{[26]}
\def\qDJMO{[2]}
\def\qEFH{[21]}
\def\qEho{[33]}
\def\qEhoPriv{[34]}
\def\qEva{[27]}
\def\qFNO{[14]}
\def\qJones{[31]}
\def\qKac{[23]}
\def\qKeMc{[4]}
\def\qKKMM{[5]}
\def\qKKMMa{[6]}
\def\qKR{[15]}
\def\qKiri{[24]}
\def\qKlMe{[11]}
\def\qKlue{[12]}
\def\qKun{[25]}
\def\qMS{[32]}
\def\qNahm{[7]}
\def\qNRT{[8]}
\def\qRav{[13]}
\def\qReck{[16]}
\def\qRC{[17]}
\def\qSchel{[19]}
\def\qTer{[9]}
\def\qVar{[22]}
\magnification=\magstep1
\parindent=0pt
\overfullrule=0pt
\def\cha{{\rm ch}}
\def\Tr{{\rm Tr}}
\def\mod{{\rm mod\ }}
\def\odd{{\rm odd}}
\def\even{{\rm even}}
\def\p{{p}}
\def\q{{p'}}
\def\chinmq{\chi^{(\p,\q)}_{n,m}(q)}
\def\chipnq{\chi^{(\p,\q)}_{\p-n,m}(q)}
\def\fullprod{\prod_{l\geq 1} (1-q^l) }
\font\gross=cmbx10 scaled\magstep3
\font\abschni=cmbx10 scaled\magstep1
\font\biggcal=cmsy10 scaled\magstep3
\font\bigcal=cmsy10 scaled\magstep1
\font\refkl=cmr8
\font\refit=cmti8
\font\refbf=cmbx8
\def\sn{\smallskip}
\def\mn{\medskip}
\def\bn{\bigskip}
\def\scs{\scriptstyle}
\def\ds{\displaystyle}
\def\W{${\cal W}$}
\setbox6=\hbox{{\bigcal W}}
\def\Wg{{\bigcal W}\hskip-\wd6\hskip 0.4 true pt{\bigcal W}}
\setbox7=\hbox{{\biggcal W}}
\def\Wgg{{\biggcal W}\hskip-\wd7\hskip 0.5 true pt{\biggcal W}}
\def\chapter#1{{\abschni #1}\mn}
\def\ChainComLi{\baselineskip20pt\lineskip3pt\lineskiplimit3pt}
\def\NormLiSkip{\baselineskip=\normalbaselineskip
\lineskip=\normallineskip\lineskiplimit=\normallineskiplimit}
\def\mapright#1{\smash{\mathop{\hbox to 50pt{\rightarrowfill}}\limits^{#1}}}
\def\mapdown#1{\Big\downarrow\rlap{$\vcenter{\hbox{$\scriptstyle#1$}}$}}
\def\ChainComplex#1{\def\normalbaselines{\ChainComLi}
\matrix{#1}\def\normalbaselines{\NormLiSkip}}
\font\HUGE=cmbx12 scaled \magstep4
\font\Huge=cmbx10 scaled \magstep4

\font\namen=cmr10 scaled \magstep1

\font\klein=cmr7 
 
\nopagenumbers
\pageno = 0
\centerline{\HUGE Universit\"at Bonn}
\vskip 10pt
\centerline{\Huge Physikalisches Institut}
\vskip 3cm
\centerline{\gross Path Spaces and {\Wgg}-Fusion in Minimal Models}
\vskip 0.4cm
\centerline{by}
\vskip 0.4cm
\centerline{\namen Johannes Kellendonk, Michael R\"osgen}
\vskip 5pt
\centerline{\namen and Raimund Varnhagen }
\vskip 2cm
\centerline{\bf Abstract}
\vskip 15pt
\noindent
Product forms of characters of Virasoro minimal models are obtained
which factorize into $(2,\odd)\times(3,\even)$ characters. 
These are related by generalized Rogers-Ramanujan identities to sum forms
allowing for a quasiparticle interpretation. 
The corresponding dilogarithm identities
are given and the factorization is used to analyse the related path space
structure as well as the fusion of the maximally extended chiral algebra.
\vfill
\centerline{\bf Int.\ J.\ Mod.\ Phys.\ A9 (1994) 1009-1023}
\vfill
\settabs \+&  \hskip 110mm & \phantom{XXXXXXXXXXX} & \cr
\+ & Post address:                       & BONN-HE-93-04   & \cr
\+ & Nu{\ss}allee 12                     & Bonn University & \cr
\+ & D-5300 Bonn 1                       & January 1993    & \cr
\+ & Germany                             & {\klein ISSN-0172-8733}& \cr
\+ & e-mail:                             &  \              & \cr
\+ & roesgen@avzw02.physik.uni-bonn.de   & hep-th/9301086  & \cr
\+ & raimund@avzw02.physik.uni-bonn.de   &  \              & \cr
\eject
\pageno=1
\footline{\hss\tenrm\folio\hss}
\chapter{1. Introduction}
Rogers-Ramanujan type identities already appeared in
the Eight-Vertex-SOS model \qABF, whose local height probabilities
can be identified \qDJMO\ with the characters of minimal models
in CFT \qBPZ.
New developments emphasize the deep meaning of the occurrence of
these identities for characters in statistical models as well
as in CFTs. The sum sides of the identities lead to a quasiparticle
interpretation \qKeMc\qKKMM\qKKMMa\ as well as to dilogarithm identities
\qNahm\qNRT\qTer\ which already appeared in integrable quantum field 
theories and statistical models \qBaRe\qKlMe\qKlue\qRav.
Their product sides encode the structure of the path spaces describing the 
combinatorics (embedding structure) of the corresponding highest weight 
representations (HWRs). 
In this article we observe that product forms of 
characters of certain sectors of the Virasoro minimal models factorize
into characters of $(2,\odd)$ and $(3,\even)$-models. In fact, 
this applies exactly to those
sectors which are invariant under the simple current, therefore belonging
to the maximally extended chiral algebra.
This is used to construct a path space generalizing
\qFNO\qKR\ where the characters of the $(2,\q)$-models were given on a
fusion graph. Moreover the fusion of 
the maximally extended chiral algebra, a \W-algebra with one extra generator,
can be obtained following the ideas in Ref.\ \qReck.
\sn
The paper is organized as follows:
In the second section we derive a factorization of modular
characters of minimal (even,odd)-models which implies the factorization of
adjacency matrices defining the corresponding path Hilbert space
in the third section.
The fourth section will elucidate that this
factorization property carries over to the fusion rules of
the corresponding \W-algebra.

\bn
\chapter{2. Factorizing Characters}
For certain HWRs of the Virasoro
minimal models we develop expressions of the modular characters
(modular forms) which are of a product form or of a sum form, corresponding
to the two sides of a generalized Rogers-Ramanujan identity.
It turns out that there exists a standard
factorization into characters of $(2,\odd)$ and $(3,\even)$-models.
\sn
We write
$c(\p,\q)=1-6{{(\p-\q)^2}\over{\p\q}}$
for the central charge and
$ {h^{(\p,\q)}_{n,m}} = { {(\p m-\q n)^2 - (\p-\q)^2} \over {4\p\q} } $
for the conformal weights for the $(n,m)$-sector of a $(\p,\q)$-model.
According to Rocha-Caridi \qRC, the embedding structure
of $(\p,\q)$-Virasoro minimal models implies that the
modular characters
$\chi (q) = q^{-c/24}\ \Tr\ q^{L_0}$, the trace running over a
HWR, are given by
$$
\chinmq =  {{q^{-c/24}} \over {\fullprod}}
          \sum_{k\in\BZ} \left( q^{a(k)} - q^{b(k)} \right)
          \eqno (2.1)
$$
with $a(k)=h^{(\p,\q)}_{n+2\p k,m}$ and
$b(k)=h^{(\p,\q)}_{n+2\p k,-m}$.
Accordingly we get
$$
q^{c/24}\chinmq\fullprod
= q^{{h^{(\p,\q)}_{n,m}}}
\sum_{k\in\BZ}
q^{\p\q k^2}\left( q^{k(\p m-\q n)}-q^{k(\p m+\q n)+nm} \right).
\eqno(2.2) 
$$
In the following, we will also make use of the shorthand notation
$$
\cha^{(\p,\q)}_{n,m}(q) =
q^{{{c(\p,\q)}\over{24}} - h^{(\p,\q)}_{n,m}} \chi^{(\p,\q)}_{n,m}(q).
$$

\mn
{\bf $\bf (2\nu ,\q)$-models:}
For $p=2\nu$ and $n=\nu$ one obtains from (2.2)
$$
\cha^{(2\nu ,\q)}_{\nu ,m}(q)
= \prod_{l\geq1} (1-q^l)^{-1}
\sum_{k\in\BZ} (-)^k q^{{\nu\q\over 2} k^2 +
{(2\nu m-\q \nu)\over 2}k},
\eqno(2.3)
$$
which by Jacobi's triple product identity 
$$
\prod_{l\geq 1} (1-v^l)(1+v^{l-{1\over 2}}w)(1+v^{l-{1\over 2}}w^{-1})
= \sum_{k\in\BZ} v^{{1\over 2}k^2}w^k
$$
for $v=q^{\nu\q}$ and $w=-q^{{(2m - \q)\nu}\over 2}$
implies
$$
\eqalignno{
\cha^{(2\nu ,\q)}_{\nu ,m}(q)
&= \prod_{l\not\equiv 0,\pm {\nu m}\ \mod{\nu\q}, l\geq 1}
(1-q^l)^{-1} &(2.4)\cr
&= \sum_{n_1 \cdots n_K \geq 0}
{{q^{N_1^2+\cdots + N_K^2 + N_{\nu m}+\cdots + N_K }}
\over{(q)_{n_1} \cdots (q)_{n_{K-1}}(q^{\gamma_\nu})_{N_K} }},&(2.5)
}$$
where $(q)_n = (1-q)\cdots(1-q^n)$,
$N_i = \sum_{j=i}^K n_j$,
$K = {{\nu\q + \gamma_\nu}\over 2} - 2$
and $\gamma_\nu = \cases{2 &for $\nu$ even\cr 1 &otherwise.\cr}$
\vskip 1mm
The equality of the product form (2.4) and the
sum form (2.5) follows from Ref.\ \qBre.
For odd $\nu$ these characters coincide with characters of
some $(2,\odd)$-model, and
$\sum_j N_j^2={\bf n}{\cal C}^{-1}_K{\bf n}$,
${\cal C}_K={(2-{\cal T_K})}$,
${\cal T}_K$ being the adjacency matrix of the tadpole graph
$A_{2K}/\BZ_2$.
\sn
It is noteworthy that the characters we find in this way are
exactly those being invariant under the simple current \qSchel.
Therefore they are also characters of the \W-algebra ${\cal W}(2,\delta)$
with $\delta = {{(\nu-1)(\q-2)}\over2}$ at the central charge given
by $c(2\nu,\q)$ \qBlu\qEFH\qVar:
$$
\chi^{{\cal W}(2,\delta)}_{\nu ,m}(q)=\chi^{(2\nu ,\q)}_{\nu ,m}(q)
\qquad\qquad
\cha^{{\cal W}(2,\delta)}_{\nu ,m}(q)=\cha^{(2\nu ,\q)}_{\nu ,m}(q) .
$$
\sn
The above form of the characters allows us to prove the {factorization} of
the $(2\nu,\q)$-characters: By (2.4) we get
$$
\cha^{(2\nu ,3)}_{\nu ,1}(q)
= \prod_{k\not\equiv 0,\pm\nu\ \mod 3\nu} (1-q^k)^{-1}
\quad {\rm and} \quad
\cha^{(2,\q)}_{1,m}(q)
= \prod_{k\not\equiv 0,\pm m\ \mod\q} (1-q^k)^{-1}
$$
and therefore
$$
\eqalign{
\cha^{(2\nu ,3)}_{\nu ,1}(q)\
&\cha^{(2,\q)}_{1,m}(q^\nu)
= \prod_{k\not\equiv0,\pm\nu\ \mod 3\nu}(1-q^k)^{-1}
  \prod_{k\not\equiv0,\pm m\ \mod \q}(1-q^{\nu k})^{-1}\cr
&= { {\ds \prod_{k\geq1} (1-q^{\nu k}) }
     \over 
     {\ds \prod_{k\geq1} (1-q^k)}     }\ 
{  {\ds \prod_{k\equiv0,\pm m\ \mod\q}(1-q^{\nu k}) }
   \over
   {\ds \prod_{k\geq1} (1-q^{\nu k}) }   }
={ {\ds \prod_{k\equiv 0,\pm\nu m\ \mod \nu\q} (1-q^k) }
   \over
   {\ds \prod_{k\geq 1} (1-q^k) } }\cr
&=\cha^{(2\nu,\q)}_{\nu ,m}(q). 
}\eqno (2.6)
$$
By uniqueness of the prefactor $q^{c/24 - h}$ making the
characters given as product expressions into modular forms \qKac,
this relation carries over to the modular characters
$$
{\chi^{(2\nu,3)}_{\nu,1}(q) \chi^{(2,\q)}_{1,m}(q^\nu)}
= \chi^{(2\nu,\q)}_{\nu,m}(q) .
$$
\sn
The sum form of the characters also allows to derive dilogarithm
identities for the effective central charge of the $(2,\odd)$-model,
following the calculations in Refs.\ \qNahm\qNRT:
The leading divergence of the characters is given by $c_{\rm eff}$
or can be calculated by a saddle point approximation.
The positive solutions of the saddle point condition
$$
1-\xi_i^{\beta_i} = \prod_{j\leq i} \xi_j^{2j} \prod_{j>i} \xi_j^{2i}
\qquad i=1,\cdots ,K \eqno(2.7)
$$
with $\beta_i = 1 + \delta_{iK} (\gamma_\nu -1)$
are given by 
$$
1-\xi_i^{\beta_i} = \left(
{\sin {\pi\over \nu\q} \over \sin{(i+1)\pi\over \nu\q}}
\right)^2 .
$$
Inserting into the Rogers dilogarithm function
$L(z):={1\over 2}\log(z)\log(1-z)-\int_0^z {\log(1-w)\over w}dw$, 
one obtains
$$
c_{\rm eff} = {1\over {L(1)}}
        \sum_{i=1}^K {1\over {\beta_i}} L(1-\xi_i^{\beta_i})
 = {1\over {2 L(1)}}
        \sum_{i=2}^{\nu\q -2} 
 L\left(\left(
{\sin {\pi \over \nu\q}\over \sin{i\pi \over \nu\q}}
\right)^2\right)                   \eqno(2.8)
$$
for the effective central charge of the
$c(2\nu,\q)$-theory under consideration.
This identity already appeared in Refs.\ \qKiri\qKun\ in the context of
$SU(2)$-WZW-models.
\mn
{\bf $\bf (3\nu,\q)$-models:} For $\p=3\nu$, using the MacDonald
identity for the Cartan matrix 
$A=\left(\matrix{2&-4\cr-1&2\cr}\right)$ \qKac\ (Watson identity)
$$
\eqalign{
&\prod_{l\geq 1} (1-u^{2l}v^{l})(1-u^{2l-1}v^{l-1})
(1-u^{2l-1}v^{l})(1-u^{4l-4}v^{2l-1})(1-u^{4l}v^{2l-1}) \cr
&= \sum_{k\in\BZ} \left(
u^{3k^2-2k}v^{(3k^2+k)/2} - u^{3k^2-4k+1}v^{(3k^2-k)/2}
\right)}
$$
with $u=q^{\nu m}$ and $v=q^{2\nu (\q-m)}$,
we obtain from (2.2) with $n=\nu$
$$
\cha^{(3\nu,\q)}_{\nu,m}(q)
= \prod_{
{\scs l\not\equiv 0, \pm {\nu m}, \pm {2\nu\q}, \pm {(2\q+m)\nu},}
\atop
{\scs \pm {2\nu (\q+m)}\  \mod{4\nu\q}, l\geq 1}
} (1-q^l)^{-1},
\eqno(2.9)
$$
proving a conjecture in Ref.\ \qChr.
In a similar way as for (2.4), a sum form of (2.9) can be written
as a product of sums of type (2.5). However, the factorization
now involves more factors than in the $(2\nu,\q)$-case above.
\mn
{\bf\W-characters:}
According to an argument given by P.~Christe \qChr, the above characters
are a complete list of Virasoro characters of minimal models
factorizing in the form
$
\prod_{l\geq 1} (1-q^l)^{n_l}
$
with exponents $n_l$ restricted to $\{0,-1\}$.
However, similar product forms can be obtained for $\cal W$-algebra
characters
which arise from summing up Virasoro characters within a simple
current orbit \qBlu\qEFH\qVar.
This might be exemplified by the case $\nu=2\mu$ in the context
of $(2\nu,\q)$-models.
Starting again from Rocha-Caridi -
$$
q^{c/24-{h^{(\p,\q)}_{n,m}}} (\chinmq - \chipnq) \fullprod
= \sum_{k\in {1\over 2}\BZ} q^{\p\q k^2} \left(
q^{(\p m-\q n)k} - q^{(\p m+\q n)k+nm}
\right)
$$

- for $\p=4\mu ,\  n=\mu$, application of the Jacobi triple product
identity yields
$$
q^{c/24-{h^{(4\mu,\q)}_{\mu,m}}}
({\chi^{(4\mu,\q)}_{\mu,m}(q)} - {\chi^{(4\mu,\q)}_{3\mu,m}(q)})
= \prod_{l\geq 1} (1-q^l)^{-1}
\prod_{l\equiv0,\pm 2m\ \mod \q} (1-q^{\mu l / 2} ).
\phantom{xxx}\eqno (2.10)
$$
For $\mu$ odd the corresponding \W -algebra ${\cal W}(2,\delta)$
with $\delta = {{(2\mu -1)(\q-2)}\over2}$ is fermionic and the \W -character
can be obtained as a sum of Virasoro characters by making use of
the symmetry properties of (2.10) under $q^{1/2} \to -q^{1/2}$:
$$
\eqalign{
&q^{c/24-{h^{(4\mu,\q)}_{\mu,m}}}\chi^{{\cal W}(2,\delta)}_{\mu,m} (q) =
q^{c/24-{h^{(4\mu,\q)}_{\mu,m}}}
({\chi^{(4\mu,\q)}_{\mu,m}(q)} + {\chi^{(4\mu,\q)}_{3\mu,m}(q)}) \cr
&= \prod_{l\not\equiv 0,\pm\mu\ \mod 3\mu} (1-q^l)^{-1}
   \prod_{l\not\equiv 0,\pm m\ \mod\q} \left( 1-q^{\mu l} \right)^{-1}
   \prod_{ {\scs l\equiv 0,\pm 2m\ \mod\q}\atop{\scs l\ \odd} }
     \left( 1+ q^{{\mu\over2}l} \right).
}
$$
According to (2.6) this can be rewritten as
$$
\cha^{{\cal W}(2,\delta)}_{\mu,m} (q) =
\cha^{(2\mu,\q)}_{\mu,m}(q) 
\prod_{ {\scs l\equiv 0,\pm 2m\ \mod\q}\atop{\scs l\ \odd} }
     \left( 1+ q^{{\mu\over2}l} \right).
\eqno(2.11)
$$

As before, (2.10) and (2.11) imply multiple sum forms, now for the
\W-characters, from which one again might derive dilogarithm
expressions describing the asymptotics.
Most presumably, similar sum and product forms do exist for more general
$\cal W$-algebra characters. We also point out that the
multiple-sum form (2.5) is by no means unique, and that there even
exist simple-sum expressions in some cases \qEva.
\bn
\chapter{3. Factorization of Path Spaces}
\def\HilbLongAB{{<{\cal S}({\cal A}({{\q-1}\over 2})
              \otimes{\cal B}(\nu),\nu ({{\q-1}\over 2}-m))>_\BC}}
\def\HilbShortAB{{{\cal H}_{\cal AB}\left( \q,m,\nu \right)}}
\def\HilbLongB{{<{\cal S}({\cal B}(\nu),0)>_\BC}}
\def\HilbShortB{{{\cal H}_{\cal B}\left( \nu \right)}}
\def\HilbLongA{{<{\cal S}({\cal A}({{\q-1}\over 2}),
               {{\q-1}\over 2}-m)>_\BC}}
\def\HilbShortA{{{\cal H}_{\cal A}\left(\q,m\right)}}
The above factorization of the characters carries over to the path space
description of the corresponding CFT. This shall be exemplified for the
$(2\nu,\q)$-models.
\sn
Consider the set of sequences $(m_j)_{j\geq 0}$
taking values in $\{ 0, \cdots , n-1\}$
which are constrained by a matrix ${\cal C}\in M_n(\{0,1\})$
and have initial value $m$
$$
{\cal S}({\cal C},m) = \left\{ (m_j)_{j\geq 0}
\in \{ 0, \cdots , n-1\}^{\BN}
\mid m_j=0
\hbox{\rm \ for }j\gg0; m_0=m, {\cal C}_{m_{j+1}+1,m_j+1} = 1
\right\}.
$$
${\cal C}$ shall be interpreted as adjacency matrix of a (labelled) graph
which we call ${\cal C}$-graph. ${\cal S}({\cal C},m)$ may then
be understood as the set
of all finite paths over this graph that start at node $m$.
${\cal H}_{\cal C}:=<{\cal S}({\cal C},m)>_\BC$
is the $\BC$-vector space freely generated by
$\cal S$ and becomes a Hilbert space by introducing the
inner product in which different paths are orthonormal.
On this we consider the endomorphism ($h\in\BQ$)
$$
L_0^{(h)} (m_j)_{j\geq 0} = ( h + \sum_{k\geq 0} k m_k )
\ (m_j)_{j\geq 0} \eqno (3.1)
$$
which is also motivated by the form of the corner transfer matrix of the
 corresponding statistical model \qBax.
Define ${\cal A}(r),{\cal B}(r)\in M_r(\{0,1\})$ by
        $({\cal B}(r))_{ij}=1$ and
$({\cal A}(r))_{ij}=
\cases{1, &if $i+j\leq r+1$\cr 0  &otherwise\cr}$ ; e.g.
$$
{\cal A} (3) = \left(\matrix{1&1&1\cr
                            1&1&0\cr
                            1&0&0\cr}\right)
\quad {\rm and} \quad
{\cal B} (3) = \left(\matrix{1&1&1\cr
                             1&1&1\cr
                             1&1&1\cr}\right)
$$
are the adjacency matrices of the graphs
in figure 1. In the following, ${\cal C}$ is a tensor product
of two matrices of the above type, and the labelling of the
corresponding graph has to be fixed by the order of the factors
(see e.g.\ figure 2).
We rewrite (hence defining this order)
$$
\eqalign{
{\cal S}&\left({\cal A}({{\q-1}\over 2})\otimes{\cal B}(\nu),\nu
({{\q-1}\over 2}-m)\right)
= \{
(l_j)_{j\geq 0} \in \BN^{\BN} \mid l_j = \nu m_j + n_j, \cr
&\phantom{\ \ \ \ \ \ \ \ \ \ }m_0 = {{\q-1}\over 2}-m,\
m_j + m_{j+1} \leq {{\q-1}\over 2}-1,\ 0\leq m_j,\ 0\leq n_j\leq \nu-1
\},\cr}
$$
and use
$$\HilbShortAB := \HilbLongAB $$
as well as
$\HilbShortA := \HilbLongA\quad{\rm and}\quad\HilbShortB := \HilbLongB $
as a shorter notation for the path Hilbert spaces.
Using the notation in the definition of ${\cal S}$,
$$
q^{L_0} (l_j) = q^{\sum_i il_i} (l_j) = q^{\sum_i i(\nu m_i + n_i)} (l_j)
= {q^\nu}^{\sum_i im_i} q^{\sum_i in_i} (l_j)
$$
implies the factorization property
$$
\Tr_\HilbShortAB\ q^{L_0^{(0)}}
= \Tr_\HilbShortA\ q^{\nu L_0^{(0)}}\
   \Tr_\HilbShortB\ q^{L_0^{(0)}} . \eqno (3.2)
$$
Due to the constraints of the sequences over the ${\cal B}$-graph
and by use of (2.4),
$$
\Tr_\HilbShortB\ q^{L_0^{(0)}}
= \prod_{i\geq 1} \left( \sum_{j=0}^{\nu -1} q^{ij}\right)
= \prod_{k\not\equiv 0,\pm\nu\ \mod 3\nu} (1-q^k)^{-1}
=\cha^{(2\nu ,3)}_{\nu ,1}(q) .\eqno (3.3)
$$
For the sequences over ${\cal A}$-graphs which appear as fusion graphs
of $(2,\odd)$-models it is known \qFNO\qKR\ that
$$
\Tr_\HilbShortA\ q^{L_0^{(0)}}
= \prod_{k\not\equiv 0,\pm m\ \mod\q} (1-q^k)^{-1}
= \cha^{(2,\q)}_{1,m}(q) .\eqno (3.4)
$$
Equations (2.6) and (3.2) therefore imply
the following result:
\sn
{\it
Assume $2\nu$, $\q$ coprime and $1\leq m\leq {{\q-1}\over 2}$.
Then the modular characters of the corresponding sectors
of Virasoro minimal models are}
$$
\chi^{(2\nu,\q)}_{\nu,m}(q) = q^{-{c(2\nu,\q)\over{24}}}
\Tr_\HilbShortAB\ q^{L_0^{(h^{(2\nu,\q)}_{\nu,m})}}.
\eqno (3.6)
$$

\mn
We briefly comment on a quasiparticle-like structure of the path
spaces.
The sum side of Rogers-Ramanujan identities has an interpretation
in terms of quasiparticles occuring in statistical models in the
sense of R.\ Kedem and B.M.\ McCoy \qKeMc.
In Refs.\ \qKKMM\qKKMMa\ the notion of quasiparticles was used for  
characters in a more general context \qTer,
and we will use this term for a similar structure
on the path spaces. 
Whereas the product side (2.4) always gives a direct description of
the structure of the path space by the constraints, 
the sum side (2.5) 
can be interpreted as the partition function of $K$ different
types of quasiparticles, which here are equivalence classes of 
elementary parts of the sequence. 
In this picture, a quasiparticle is excited
by moving the corresponding patterns to
the right along the sequence according to the constraints
of the path space.
Particles of different types can be excited 
independently, whereas particles of one type have minimal 
distance two. The latter condition originates from the
structure of the annihilating ideal \qFNO\ and might be regarded as
a generalized
Pauli principle. In (2.5) the value of the summation 
variable $n_j$ is the number of the quasiparticles of 
type $j$, the factor $1/(q)_{n_j}$ generating all its 
excitations.
\sn
The ground state of the $(n_1,\cdots,n_K)$-particle sector
is ordered with the energy weight (the number of the
particle type) of the particles decreasing 
from the left to the right and therefore is given by the 
$q$-exponent in the denominator of (2.5) where the missing 
linear terms encode the path initial conditions. 
\sn
In the case $\nu=1$, all ${\q-3}\over2$ quasiparticles occuring
are of the form $(m_l)=(\cdots k,j-k\cdots)$,\   
$j=1,\cdots,{{\q-3}\over2}$ being the energy weight (and denoting
the particle type), and
$k=1,\cdots,j$ together with the position counting different energy 
configurations within one type (pattern equivalence class).
In general, there are $N:={{\q-1}\over2}\nu-1$
(the number of allowed nonzero values on the paths) such simple
quasiparticles as $(\cdots k,j-k\cdots)$\ $(j=1,\cdots,N)$ and
$K-N=\left[{\nu\over2}\right]$ additional ones.
\sn
In the simplest case, $\nu=2$, the one additional 
quasiparticle is of the form $(\cdots N-k,1+k\cdots)$, 
where according to the constraints by the adjacency matrix 
only even values of $k=0,2,\cdots,N-1$ are allowed.
Hence particles of this type can only be moved in energy 
steps of two. This is the origin of the $\gamma_\nu=2$ 
exponent in the $K^{\rm th}$ $q$-bracket in (2.5).
\mn
It is evident that the product expression of the fermionic
{\W-characters} in (2.11)
implies similar path spaces. In fact, for $\mu$ odd the energy
operator may be taken to be
$$
L_0^{(h)} (l_j)_{j\geq 0} =
( h + \sum_{k\geq 0} {k\over 2} l_k )\ (l_j)_{j\geq 0}
\eqno (3.7)
$$
where the sequences $(l_i)_{i\geq 1}$ are subject to the conditions
$(l_{2i})_{i\geq 1}\in
{\cal S}({\cal A}({{\q-1}\over2})\otimes{\cal B}(\mu))$
and $l_i \in \{ 0 , \mu \}$ for $i$ odd with $i\equiv 0,\pm 2m\ \mod \q$,
finally $l_i=0$ for the remaining positions.
The \W-algebra characters which are already contained in the 
result (3.6) might be described in a way close to above, 
if one again uses $L_0$ of the form (3.7), 
restricting to the even positions
of the sequence; the rules of ${\cal A}\otimes{\cal B}$ then
describe the matching of sequence positions of distance two.
The odd positions are simply set to zero in this case.
In this picture, we see that the corresponding path spaces
on Bratteli-like diagrams (figure 3) do not have the same 
structure  for all sectors, but a projection to paths over 
graphs is not obvious in the case of summed characters.
\bn
\chapter{4. Factorization of {\Wg}-Fusion}
Recall that the Virasoro characters which factorize according to
(2.6) are exactly those which are also characters of the
maximally 
extended chiral algebra (the corresponding \W-algebra).
This suggests that the path space structure examined above is
related to the extended model rather than to the Virasoro
minimal model. 
In particular, we expect this structure to contain the fusion of the 
corresponding \W-algebra, since it
governs the relative size of the sectors, i.e.\
the quantum dimensions.
\sn
In the following we refer to methods developed by
A.~Recknagel \qReck. This approach is motivated by
algebraic field theory
and relates the fusion to 
the $K$-theory of the algebra of observables or - simplifying -
to some "good" approximation of it. 
Recall that one can describe the $K_0$-group of an 
associative algebra
as the abelian group associated to the semi-group of stable
equivalence classes of idempotents in the matrices having entries
in that algebra \qBla. 
\sn
Having established a grading preserving isomorphism between the
HWRs of a $(2\nu,\q)$-model which
correspond to \W-sectors
and the path spaces of the ${\cal A}({{\q-1}\over{2}})\otimes
{\cal B}(\nu)$-graph,
one might expect that the $AF$-algebra which
is defined by the path space (the path space being understood as 
the Bratteli diagram of this algebra)
already carries much of the structure of the CFT. Of course the 
general situation is not as convincing as for $\nu=1$ \qFNO\qKR,
since first, we do not have the same graph for all sectors, and second,
we cannot relate the path space directly to an annihilating ideal. 
Note also that for $\nu>1$ the ${\cal A}({{\q-1}\over{2}})\otimes
{\cal B}(\nu)$-graph cannot be a fusion graph.
But still, at least for $\nu = 2$ the
corresponding $AF$-algebra contains the
fusion of the ${\cal W}(2,{{\q-2}\over{2}})$
in the sense expressed in Ref.\ \qReck: 
Taken for granted that it is a good enough
approximation of the observable algebra, the fusion ring should be a maximal
commutative subring of
the ring of positive endomorphisms of its ordered $K_0$-group \qBla.
The question whether, given such a subring, any set of generators
(generating the subring as a $\BZ$-module)
satisfying the fusion rule axioms determines the same fusion rule,
is not yet completely analysed, but in many cases the choice
of one generator seems to be sufficient to fix it.

The directed system of $K_0$-groups determined by the path space
${\cal S}({\cal A}({{\q-1}\over{2}})\otimes {\cal B}(\nu))$ 
(the starting node only influences its beginning but not the algebraic
limit) is
$$
 \def\einbet{\mapright{{\cal A}({{\q-1}\over{2}})\otimes{\cal B}(\nu)}}
 \def\bzrs{\BZ^{{\q-1}\over{2}}\otimes\BZ^\nu}
 \ChainComplex{
   \cdots&\einbet&\bzrs&\einbet&\bzrs&\einbet&\cdots\cr
   }.  \eqno (4.1)
$$
The factorization of the graph suggests to
first consider each graph separately and then
to tensorize the resulting endomorphisms.
The first graph already occurred in Ref.\ \qKR\
as the fusion graph of the
$h_{min}$-field of the $(2,\q)$-Virasoro model.
Here the adjacency matrix is bijective so that it may be used to
identify the $K_0$-group of the algebraic limit of (4.1) 
with $\BZ^{{\q-1}\over{2}}$.
A maximal commuting subset of all positive $K_0$-endomorphisms is 
linearly generated by 
$p_k({\cal T}_{{\q-1}\over{2}})$, $k=0,\cdots,{{\q-3}\over2}$, the
Chebychev polynomials of the second kind (defined by $p_0=1$,
$p_1(x)=x$, $p_{n+2}(x)=xp_{n+1}(x)-p_n(x)$) evaluated at
the adjacency matrix ${\cal T}_{{\q-1}\over2}$ of the
tadpole graph $A_{\q-1}/\BZ_2$ \qCas.
Moreover these polynomials are generators of the fusion ring of the
$(2,\q)$-Virasoro minimal model.
Questions of uniqueness parallel those for the $SU(2)$-WZW models which
are treated in Ref.\ \qReck\ as an example.
Concerning the second graph let us first consider the case $\nu=2$.
The ${\cal B}(2)$-graph is well known to describe the embedding of the
Majorana algebra (the even part of the Clifford algebra) of $2n$ generators
into the one with $2(n+1)$ generators. The $AF$-algebra described by the
corresponding Bratteli diagram
(the path space of the ${\cal B}(2)$-graph which may also be obtained
by the tower construction of V.~Jones \qJones)
already appeared in Ref.\ \qMS,
a suitable closure of it being the Neveu-Schwarz component of the
observable algebra of the Ising model.
Its $K_0$-group equals the dyadics $\BZ [{{1}\over{2}}]$
with the usual order relation (from $\BR$).
To obtain the fusion ring here, it is necessary to use the refined version
of the above method, as any endomorphism of $\BZ [{{1}\over{2}}]$ is given
by multiplication
with a dyadic number. These difficulties arise from the degeneracy
of the adjacency matrix ${\cal B}(2)$.
For this reason it is proposed in Ref.\ \qReck\ to
consider stationary systems of filtration preserving endomorphisms
of the directed system itself (which also covers the result for bijective
adjacency matrices). These are endomorphisms $\{\rho_n\}$
$$
\ChainComplex{
  \cdots&\mapright{{\cal B}(2)}&\BZ^2&\mapright{{\cal B}(2)}&
  \BZ^2&\mapright{{\cal B}(2)}&\cdots\cr
  &&\mapdown{\rho_n}&&\mapdown{\rho_{n+1}}&&\cr
  \cdots&\mapright{{\cal B}(2)}&\BZ^2&\mapright{{\cal B}(2)}&
  \BZ^2&\mapright{{\cal B}(2)}&\cdots\cr
}
$$
where $\rho_n=\rho$ for all $n$, the $\rho$'s being endomorphisms of $\BZ^2$
that commute with
${\cal B}(2)$ and preserve the cone defined by $z_1+z_2 > 0$.
In fact, since the energy grading may not be well defined for 
equivalence classes of Bratteli diagrams, but only for distinguished
representatives (the path spaces), not only the
direct limit but the directed system itself should encode the fusion.
It is clear that a commutative subring is now generated (as a $\BZ$-module)
by two elements.
Next to the $2\times 2$ unit matrix,
the only possible choice for the second generator which is
compatible with the axioms for a fusion ring is
$
\left(
\matrix{0&1\cr 1&0\cr}
\right),
$
the ${\cal B}(2)$-graph therefore describing a $\BZ_2$-fusion.

For arbritray $\nu$ a maximal commuting subring of endomorphisms of $\BZ^\nu$
commuting with ${\cal B}(\nu)$ is generated 
by $\omega^l$, $0\leq l<\nu$ ($\omega^\nu=id$), where $\omega$ is a cyclic
permutation of the base.
Hence, neglecting questions of uniqueness of the above mentioned type,
the ${\cal B}(\nu)$-graph yields a $\BZ_\nu$-fusion.
\sn
Although the above arguments are not yet at a rigorous level
they encourage us to conjecture 
\sn
{\it the generators of the fusion ring of
${\cal W}(2,\delta)$, $\delta={{(\nu-1)(\q-2)}\over2}$ at $c(2\nu,\q)$
to be represented by
$$\phi_{kn} := p_{k-1}{\left({\cal T}_{{\q-1}\over2}\right)}\otimes\omega^{n-1}
\qquad\qquad k=1,\cdots ,{{\q-1}\over2}\quad n=1,\cdots,\nu \eqno(4.2)$$
where $\omega$ is a generator of $\BZ_\nu$, i.e.\ by 
``$(2,\q){\rm -fusion}\otimes\BZ_{\nu}{\rm -fusion}$''.}
\sn
In fact, for the fermionic
\W-algebras ${\cal W}(2,{{\q-2}\over{2}})$, $\q$ odd,
for which some of the \W-characters already appear 
as Virasoro characters in
$(4,\q)$-models $(\nu = 2)$ the conjecture is supported 
for the Neveu-Schwarz sector by earlier results of Ref.\ \qVar\  
and in total by very recent results
of W.~Eholzer \qEho\ who derived them starting
from the representation theory of the modular group.
An explicit example is given by the 
${\cal W}(2,{3\over2})$-fusion
at c(4,5)=7/10 in figure 4.
Furthermore,
by comparison with further examples \qEhoPriv\
the decomposition of the ${\cal B}(\nu)$-matrix into 
a $\BZ_\nu$ fusion is also 
suggested  in the bosonic case.
In fact, the fusion of the untwisted sector
of ${\cal W}(2,3)$ at $c(5,6) = 4/5$, where the
sectors $h=2/3$ and $h=1/15$ are doubled due to
non vanishing $W_0$ eigenvalue, is given
by $(2,5){\rm-fusion}\otimes\BZ_3$.

\bn
\chapter{5. Conclusions and Outlook}
For all Virasoro minimal models of type $(2\nu,\q)$
we found factorizing product expressions for the characters
of the simple current invariant sectors.
In particular all unitary models are covered.
For the corresponding fermionic ${\cal W}(2,\delta)$
even more characters could be shown to factorize, in fact for 
$\nu=2$, all of them.
These are all chiral extensions of the Virasoro minimal
models by one generator of vanishing self-coupling. 
\sn
We derived a path space realization of the
corresponding HWRs, in which the factorization became manifest
in the graph adjacency. $K$-theoretic arguments were applied 
to demonstrate how the factorization carries over to the \W-fusion.
The case of $\nu=1$ is exceptional in the sense that, by lack
of a simple current, the Virasoro fusion is reproduced.
For the $\nu=2$ case, the results are established and
in total coincidence with those of Ref.\ \qEho.
By investigating a bosonic example, we checked
that this result most presumably
generalizes to all \W-algebras extending
Virasoro minimal models.
\sn
It remains to find product and sum forms for the other 
characters of these models,
and to understand in more detail the deviations in the 
structure of the path spaces, which
seem to be necessary to avoid quantum dimensions smaller
than one.
For a complete description of minimal models
the $(\odd,\odd)$-models should also
be included. They in fact show a similar factorization
structure involving more factors but again of the types
${\cal A}$ and ${\cal B}$.
\sn
We emphasize that by linear combination of \W-characters, 
one can also derive sum forms of the
Virasoro characters of minimal models, leading
to dilogarithm identities. Furthermore the
path space structure in these cases should give
a direct access to the annihilating ideal,
paving the way to generalize Ref.\ \qFNO.

\bn\bn
\chapter{Acknowledgements}
We are indebted to A.~Recknagel, who suggested to analyse the relation
of the Virasoro HWRs to \W-fusion, and to W.~Eholzer, who calculated
\W-fusion rules with different methods.
We thank both of them as well as W. Nahm and M. Terhoeven
for helpful discussions and comments.
We also thank M. Flohr for some
computational help in the ${\cal W}(2,3)$ case
as well as R. Blumenhagen, A. Honecker, R. H\"ubel and
S. Mallwitz for discussions on related subjects.

M. R\"osgen is supported by Cusanuswerk, R. Varnhagen
by Deutsche Forschungsgemeinschaft (DFG).

\def\CMP#1{{ Commun.\ Math.\ Phys.\ {\refbf #1}}}
\def\IMPA#1{{ Int.\ J.\ Mod.\ Phys.\ {\refbf A#1}}}
\def\IMPB#1{{ Int.\ J.\ Mod.\ Phys.\ {\refbf B#1}}}

\def\JSP#1{{ J.\ Stat.\ Phys.\ {\refbf #1}}}

\def\MPL#1{{ Mod.\ Phys.\ Lett.\ {\refbf #1}}}

\def\NPB#1{{ Nucl.\ Phys.\ {\refbf B#1}}}

\def\PLB#1{{ Phys.\ Lett.\ {\refbf #1B}}}

\def\PR#1{{ Phys.\ Rev.\ {\refbf #1}}}

\bn
\chapter{References}
\refkl
\parindent=12pt
\item{\qABF} G.E.~Andrews, R.J.~Baxter, P.J.~Forrester,
  {\refit Eight-vertex SOS model and generalized Rogers-Ramanujan type
  identities,\/}
  \JSP{35} (1984) 193-266
\item{\qDJMO} E.~Date, M.~Jimbo, T.~Miwa, M.~Okado,
  {\refit Automorphic properties of local height probabilities for integrable
  solid-on-solid models,\/}
  \PR{B 35} (1987), 2105-2107;
  E.~Date, M.~Jimbo, A.~Kuniba, T.~Miwa, M.~Okado,
  {\refit Exactly solvable SOS models: Local height probabilities and
  theta function identities,\/}
  \NPB{290} (1987) 231-273
\item{\qBPZ} A.A.~Belavin, A.M.~Polykov, A.B.~Zamolodchikov,
   {\refit Infinite Conformal Symmetry in Two-Dimensional
    Quantum Field Theory,\/}
  \NPB{241} (1984) 333-380
\item{\qKeMc} R.~Kedem, B.M.~McCoy,
   {\refit Construction of Modular Branching Functions from
   Bethe's Equations in the 3-state Potts Chain,\/}
   ITP-SB-92-56, hepth 9210129
\item{\qKKMM} R.~Kedem, T.R.~Klassen, B.M.~McCoy, E.~Melzer,
   {\refit Fermionic quasiparticle representations for characters of
   $G^{(1)}_1 \times G^{(1)}_1 / G^{(1)}_2$ ,\/}
   ITP-SB-92-64 / RU-92-51 / hep-th 9211102
\item{\qKKMMa} R.~Kedem, T.R.~Klassen, B.M.~McCoy, E.~Melzer,
   {\refit Fermionic Sum Representations for Conformal Field
   Theory Characters,\/}
   ITP-SB-93-05 / RU-93-01 / hep-th 9301046
\item{\qNahm} W.~Nahm, Talk given at the Isaac Newton Institute, 
  Cambridge, Sept. 1992; to appear in the proceedings of the NATO
  workshop on {\refit Low dimensional Topology and Quantum Field
  Theory\/}
\item{\qNRT} W.~Nahm, A.~Recknagel, M.~Terhoeven,
  {\refit Dilogarithm Identities in Conformal Field Theory,\/}
  BONN-HE-92-35, hepth 9211034, to appear in \IMPA{}
\item{\qTer} M.~Terhoeven,
  {\refit Lift of dilogarithm to partition identities,\/}
  BONN-HE-92-36, hepth 9211120
\item{\qBaRe} V.V.~Bazhanov, N.Yu. Reshetikhin, 
     {\refit Critical RSOS Models and Conformal Field Theory,\/}
     \IMPA{4} (1989) 115
\item{\qKlMe} T.R.~Klassen, E.~Melzer, 
     {\refit Purely elastic scattering theories and their ultraviolet
     limits,\/}
     \NPB{338} (1990) 485-528;
     {\refit Spectral flow between conformal field theories in
      (1+1)-dimensions,\/}
     \NPB{370} (1992) 511-550
\item{\qKlue} A.~Kl\"umper, P.A.~Pearce,
     {\refit Analytic Calculation of Scaling Dimensions: Tricritical
     Hard Squares and Critical Hard Hexagons,\/}
     \JSP{64} (1991) 13-76
\item{\qRav} F.~Ravanini,
     {\refit Thermodynamic Bethe ansatz for $G_k\times G_l/G_{k+l}$ coset  
     models perturbed by their $\phi_{1,1,Adj}$ operator,\/}
     \PLB{282} (1992) 73-79
\item{\qFNO} B.L.~Feigin, T.~Nakanishi, H.~Ooguri,
  {\refit The annihilating ideals of minimal models,\/}
  \IMPA{7} Suppl.\ {\refbf 1A} (1992) 217-238
\item{\qKR} J.~Kellendonk, A.~Recknagel,
  {\refit Virasoro Representations on Fusion Graphs,\/}
  \PLB{298} (1993) 329-334
\item{\qReck} A.~Recknagel,
  {\refit Fusion Rules from Algebraic $K$-Theory,\/}
  BONN-HE-92-06, to appear in \IMPA{}
\item{\qRC} A.~Rocha-Caridi,
  {\refit Vacuum Vector Representations of the Virasoro Algebra,\/}
  in {\refit Vertex Operators in Mathematics
  and Physics,\/}
  MSRI Publications \# 3 (Springer, Heidelberg, 1984) 451-473
\item{\qBre} D.M.~Bressoud,
  {\refit Analytic and combinatorial generalizations of the
  Rogers-Ramanujan identities,\/}
  Memoirs of the Am.\ Math.\ Soc.\ No.\ 227, 24 (1980)
\item{\qSchel}  A.N.~Schellekens, S.~Yankielowicz,
  {\refit Extended Chiral Algebras and Modular Invariant
  Partition Functions,\/}
  \NPB{327} (1989) 673-703
\item{\qBlu} R.~Blumenhagen, M.~Flohr, A.~Kliem, W.~Nahm, A.~Recknagel,
  R.~Varnhagen,
  {\refit $\cal W$-Algebras with Two and Three Generators,\/}
  \NPB{361} (1991) 255-289
\item{\qEFH} W.~Eholzer, M.~Flohr, A.~Honecker, R.~H\"ubel, W.~Nahm,
  R.~Varnhagen,
  {\refit Representations of $\cal W$-Algebras with Two Generators
  and New Rational Models,\/}
  \NPB{383} (1992) 249-288
\item{\qVar} R.~Varnhagen,
  {\refit Characters and Representations of New Fermionic $\cal 
  W$-Algebras,\/}
  \PLB{275} (1992) 87-92
\item{\qKac} V.G.~Kac,
 {\refit Infinite dimensional Lie Algebras,\/}
 3rd edition, (Cambridge University Press, Cambridge, 1990)
\item{\qKiri} A.N.~Kirillov, 
    {\refit Identities for the Rogers dilogarithm function connected with
    simple Lie algebras,\/}
    J.\ Sov.\ Math.\ {\refbf 52} (1989) 2450-2459,
    (Zap.\ Nauch.\ Semin.\ LOMI {\refbf 164} (1987) 121-133)
\item{\qKun} A.~Kuniba, T.~Nakanishi,
   {\refit Spectra in conformal field theories from the Rogers 
    dilogarithm,\/}
   \MPL{A7} (1992) 3487-3494; 
   {\refit Rogers dilogarithm appearing in integrable systems,\/}
   HUTP-92/A046 
\item{\qChr} P.~Christe,
  {\refit Factorized Characters and Form Factors of
  Descendant Operators in Perturbed Conformal Systems,\/}
  \IMPA{29} (1991) 5271-5286
\item{\qEva} D.E.~Evans,
  {\refit $C^*$-Algebraic Methods in Statistical Mechanics and Field 
  Theory,\/}
  \IMPB{4} (1990) 1069-1118
\item{\qBax} R.J.~Baxter,
  {\refit Exactly Solved Models In Statistical  Mechanics,\/}
  Academic Press. London, 1982
\item{\qBla} B.~Blackadar,
    {\refit $K$-Theory of Operator Algebras,\/}
    MSRI Publications \# 5, Springer, New York, 1986
\item{\qCas} M.~Caselle, G.~Ponzano, F.~Ravanini,
  {\refit Towards a classification of fusion rule algebras in rational
  conformal field theories,\/}
  \IMPB{6} (1992) 2075-2090
\item{\qJones} F.M.~Goodman, P.~de la Harpe, V.F.R.~Jones,
    {\refit Coxeter Graphs and Towers of Algebras,\/}
    MSRI Publications \# 14, Springer, New York, 1989
\item{\qMS} G.~Mack, V.~Schomerus,
  {\refit Conformal Field Algebras with Quantum Symmetry from the Theory
   of Superselection Sectors,\/}
  \CMP{134} (1990) 139-196
\item{\qEho} W.~Eholzer, R.~H\"ubel,
  {\refit Fusion Algebras of Fermionic Rational Conformal Field Theories
  via Generalized Verlinde Formula,\/}
  BONN-HE-93-05
\item{\qEhoPriv} W.~Eholzer, private communication
\vfill\eject
\message{ -> 4 Postscript figures pswf[1-4].ps have to be printed manually <- }
\bye